\documentclass[manuscript]{aastex}

\slugcomment{To be published, Astrophysical Journal}

\shorttitle{RESIK observations of the solar flare X-ray continuum}
\shortauthors{Phillips et al.}

\begin{document}

\title{THE SOLAR X-RAY CONTINUUM MEASURED BY RESIK}

\author{K. J. H. Phillips\altaffilmark{1}}
\affil{Mullard Space Science Laboratory, University College London, Holmbury St Mary, Dorking,
Surrey RH5 6NT, U.K.}
\email{kjhp@mssl.ucl.ac.uk}

\and

\author{J. Sylwester and B. Sylwester\altaffilmark{2} }
\affil{Space Research Centre, Polish Academy of Sciences, 51-622, Kopernika~11, Wroc{\l}aw, Poland}
\email{js@cbk.pan.wroc.pl, bs@cbk.pan.wroc.pl}

\and

\author{V. D. Kuznetsov\altaffilmark{3} }
\affil{Institute of Terrestrial Magnetism and Radiowave Propagation (IZMIRAN), Troitsk, Moscow, Russia}
\email{kvd@izmiran.ru}

\begin{abstract}
The solar X-ray continuum emission at five wavelengths between 3.495~\AA\ and 4.220~\AA\ for 19 flares in a seven-month period in 2002--2003 was observed by the RESIK crystal spectrometer on {\it CORONAS-F}. In this wavelength region, free--free and free--bound emissions have comparable fluxes. With a pulse-height analyzer having settings close to optimal, the fluorescence background was removed so that RESIK measured true solar continuum in these bands with an uncertainty in the absolute calibration of $\pm 20$\%. With an isothermal assumption, and temperature and emission measure derived from the ratio of the two {\it GOES} channels, the observed continuum emission normalized to an emission measure of $10^{48}$~cm$^{-3}$ was compared with theoretical continua using the {\sc chianti} atomic code. The accuracy of the RESIK measurements allows photospheric and coronal abundance sets, important for the free--bound continuum, to be discriminated. It is found that there is agreement to about 25\% of the measured continua with those calculated from {\sc chianti} assuming coronal abundances in which  Mg, Si, Fe abundances are four times photospheric.
\end{abstract}


\section{INTRODUCTION}\label{intro}

The RESIK (REntgenovsky Spektrometr s Izognutymi Kristalami) X-ray spectrometer on the {\it CORONAS-F} solar spacecraft obtained numerous solar X-ray spectra in the range 3.40--6.05~\AA\ from shortly after the spacecraft launch (on 2001 July~31) to 2003 May. The instrument \citep{syl05} was a bent crystal spectrometer with four channels, the solar X-ray emission being diffracted by silicon and quartz crystals. Pulse-height analyzers enabled solar photons to be distinguished from those produced by fluorescence of the crystal material (photon energies of $\gtrsim 1.84$~keV, the K-shell ionization energy of Si) through the different energies of the photons. This has meant that the instrumental background, which has been a significant problem for most previous spacecraft crystal spectrometers (e.g. \cite{cul91}), could be eliminated for channels 1 and 2 (Si~111 crystal, $2d = 6.27$~\AA) since for these channels the range of solar photon energies are substantially different (3.26--3.64~keV and 2.90--3.24~keV). For channels 3 and 4 (quartz~$10\bar 10$ crystal, $2d = 8.51$~\AA), the photon energies are closer and the discrimination is only partial, but the fluorescence can nevertheless be accurately estimated. Thus, a means of measuring the solar continuum is offered. The spectral ranges of RESIK for on-axis solar sources are: channel~1, 3.40--3.80~\AA; channel~2, 3.83--4.27~\AA; channel~3, 4.35--4.86~\AA; and channel~4, 5.00--6.05~\AA. RESIK was uncollimated to maximize the instrument's sensitivity; this leads to some degree of spectral confusion only on the rare occasions when two simultaneous flares occurred on the Sun. While previous work on RESIK spectra has concentrated on line spectra \citep{syl06, phi06a, chi07,syl08,syl09}, here we discuss continuum emission observed during flares. There are several portions of the X-ray spectrum observed by RESIK in channels~1 and 2 that are practically free of lines and therefore enable the X-ray continuum flux to be estimated. Few such observations have provided accurate continuum fluxes in this wavelength region.

With estimates of temperature $T_{\rm GOES}$ and emission measure of the emitting regions available from the flux ratio of the two channels of {\it GOES}, we have been able to examine the continuum emission at available wavelengths as a function of $T_{\rm GOES}$. We have compared this with calculated continua from the {\sc chianti} atomic database and code \citep{lan06}.  The wavelength range concerned (3.5--4.2~\AA) is of particular interest because within it free--bound emission is comparable to free--free emission, so the accuracy of particularly the free--bound emission calculations can be verified. The free--bound emission depends on the set of abundances used in the calculations, a coronal set (\cite{fel92, flu99}) giving rise to greater emission than a photospheric set  \citep{gre07,asp09}. The accuracy of the photometric calibration of RESIK is such that this difference can be detected. Further, recent calculations \citep{bro09} suggest that free--bound emission may sometimes significantly contribute to the total non-thermal continuum during solar flare impulsive stages  and so a check on the thermal continuum at high temperatures as calculated by {\sc chianti} and other codes is very desirable.

\section{THEORETICAL CONTINUA}\label{theor_contm}

We first describe calculations using {\sc chianti} of free--free, free--bound, and two-photon continua emitted by a solar coronal plasma. These were obtained from functions available in the IDL-based SolarSoft system \citep{fre98}. The {\sc chianti} free--free continua are based on fitting formulas given by \cite{sut98} and \cite{ito00}. Ionization fractions which are needed as input to both free--free and free--bound continua were from the recent work of \cite{bry09}. Element abundances also affect the free--bound continuum. It is found that the continua in the RESIK wavelength range are made up of free--free and free--bound continua in comparable amounts. For the temperatures and wavelengths considered here, two-photon continua (arising from the de-excitation of metastable levels in H-like and He-like ions) are a factor $\sim 30$ less than either free--bound or free--free continuum emission, and were therefore neglected.

For the wavelengths considered here (3.40--4.27~\AA), free--bound emission is especially important for coronal abundances at flare-like temperatures ($\lesssim 20$~MK). Large contributions to the total emission are made by recombination to Si, Fe, and Mg ions, and to some extent O ions. The coronal abundances of Si, Fe, and Mg are higher than the corresponding photospheric abundances according to the ``FIP" effect, for which elements with first ionization potential (FIP) lower than 10~eV have enhanced coronal abundances. The free--bound emission is thus greater for coronal abundances by amounts that depend on the exact abundance set assumed. In this work, we took coronal abundances from \cite{fel92} and \cite{flu99}, and photospheric abundances from \cite{gre07}. (The coronal abundances of \cite{fel00} are similar to those of \cite{fel92} for C, N, O, Ne, and Ar, but similar to those of \cite{flu99} for Mg, Si, S, Ca, and Fe; the photospheric abundances of \cite{asp09} are within 20\% of those of \cite{gre07} except for Ar for which \cite{asp09} is 65\% larger.) Edge features in the continuum are formed when free electrons recombine with fully stripped and H-like Si ions at 5.08~\AA\ and 4.64~\AA\ (2.44~keV, 2.67~keV); with fully stripped and H-like Mg ions at 7.05~\AA, 6.33~\AA\ (1.76~keV, 1.96~keV); and with fully stripped and H-like S ions at 3.55~\AA, 3.85~\AA\ (3.49~keV, 3.22~keV). Recombination of electrons with a range of Fe ions gives rise to edges between 6~\AA\ and 9.1~\AA\ (1.36--2.1 keV). There are upward jumps in the free--bound emission below each of these wavelengths. This gives rise to an accumulation of emission at wavelengths $\lesssim 9$~\AA\ for a large temperature range. The S edges fall within the range considered here and so may be important. To the free--bound emission must be added free--free continuum, due mostly to H and He, with very small contributions from heavier elements.

\begin{figure}
\epsscale{1.}
\plotone{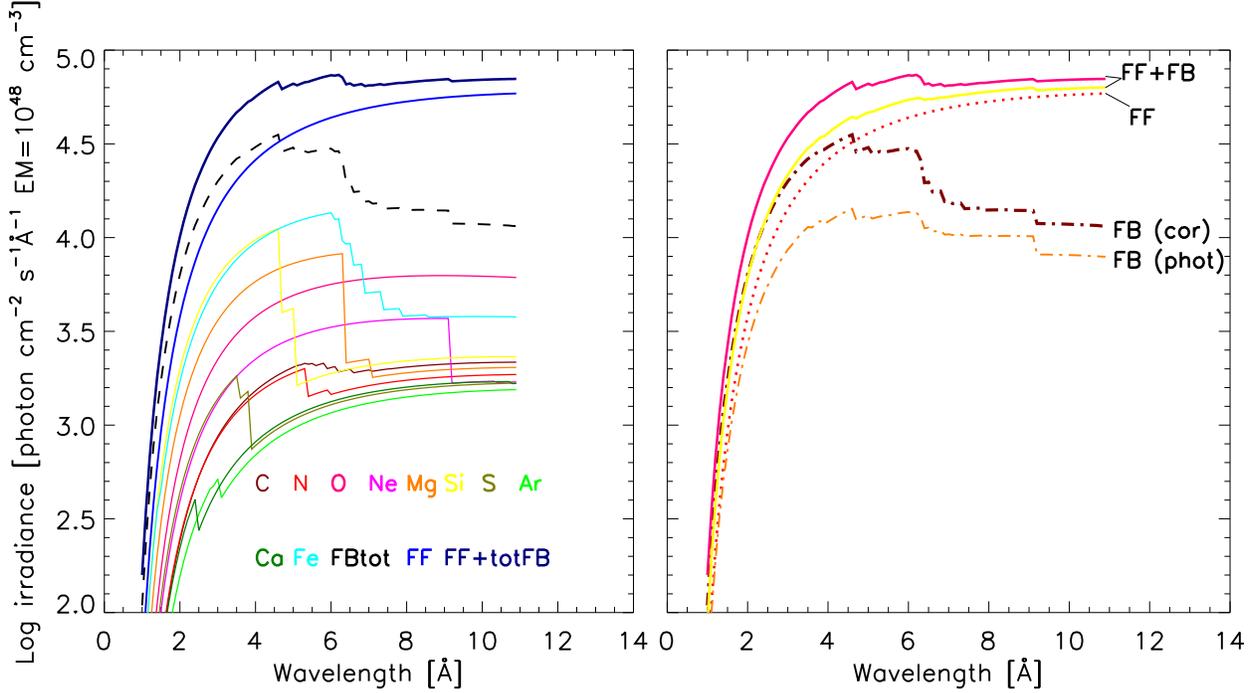}
\caption{{\it Left panel:} Continuum fluxes (irradiances, photon units) calculated from the CHIANTI atomic package at $T_e = 15$~MK and an emission measure $10^{48}$~cm$^{-3}$ plotted against wavelength (1--11~\AA). The solid line shows the sum of free--free (FF) and free--bound (FB) continua, the dotted lines the free-free and free--bound continua for all elements. Other curves show the contributions to the free--bound continua made by individual elements (O, Fe, Si, Mg, S) with coronal abundances (\cite{fel92}).  {\it Right panel:} Continuum fluxes compared for coronal (\cite{fel92}) and photospheric (\cite{gre07}) abundances for $T_e = 15$~MK. Solid lines are the total of free--free and free-bound, the dot--dash curves are for free--bound (FB), coronal and photospheric abundances indicated. The dotted curve is the free--free continuum for coronal abundances (free-free emission for photospheric abundances is about 10\% lower). [A color version of this figure is available in the on-line version of the journal, showing free--bound emission from more elements: color key is C (maroon); N (red); O (pink); Ne (magenta); Mg (orange); Si (yellow); S (olive); Ar (green); Ca (dark green); Fe (cyan); free--free emission (blue); and total emission (dark blue).] } \label{cont_diffelems}
\end{figure}

Figure~\ref{cont_diffelems} ({\it left panel}) shows the total free--free, free--bound continuum emission from {\sc chianti} and their sum in the 1--11~\AA\ range (which therefore includes the range of all RESIK detectors) for electron temperature $T_e = 15$~MK and the \cite{fel92} coronal abundance set. A volume emission measure of $10^{48}$~cm$^{-3}$ is assumed. For other temperatures, the relative continuum fluxes are quite similar. Individual contributions made by O, Si, Mg, S, and Fe ions to the free--bound emission are indicated by the different line symbols: it can be seen how large these contributions are at various wavelengths. For the 3.5--4.2~\AA\ range, the contributions to the total free--bound continuum made by Fe and Si are each 31\%,  Mg 18\%, and O 13\%. The use of coronal or photospheric abundances for these elements is clearly a matter of some importance. According to \cite{fel92}, the coronal abundances are greater than those of \cite{gre07} by factors of 1.7 (O), 4 (Mg and Si),  and 4.5 (Fe). The corresponding factors for the coronal abundance set of \cite{flu99} are 1.2 (O) and between 2.3 and 2.4 for Mg, Si, and Fe. For S, the coronal abundances are only slightly higher (between 17\% and 34\%) than photospheric abundances. Figure~\ref{cont_diffelems} ({\it right panel}) shows the differences for coronal \citep{fel92} and photospheric \citep{gre07} abundance sets. At 4.0~\AA\ the free--bound continuum is a factor $>2$ higher for coronal abundances, and the total continuum is higher by 70\%. (The free--free continuum is only $\sim 10$\% lower for the photospheric abundances of \cite{gre07}.) This is larger than the expected uncertainties (20\%) in the RESIK absolute flux calibration, offering the possibility of distinguishing element abundance sets through RESIK observations of the continuum emission.

\begin{figure}
\epsscale{.9}
\plotone{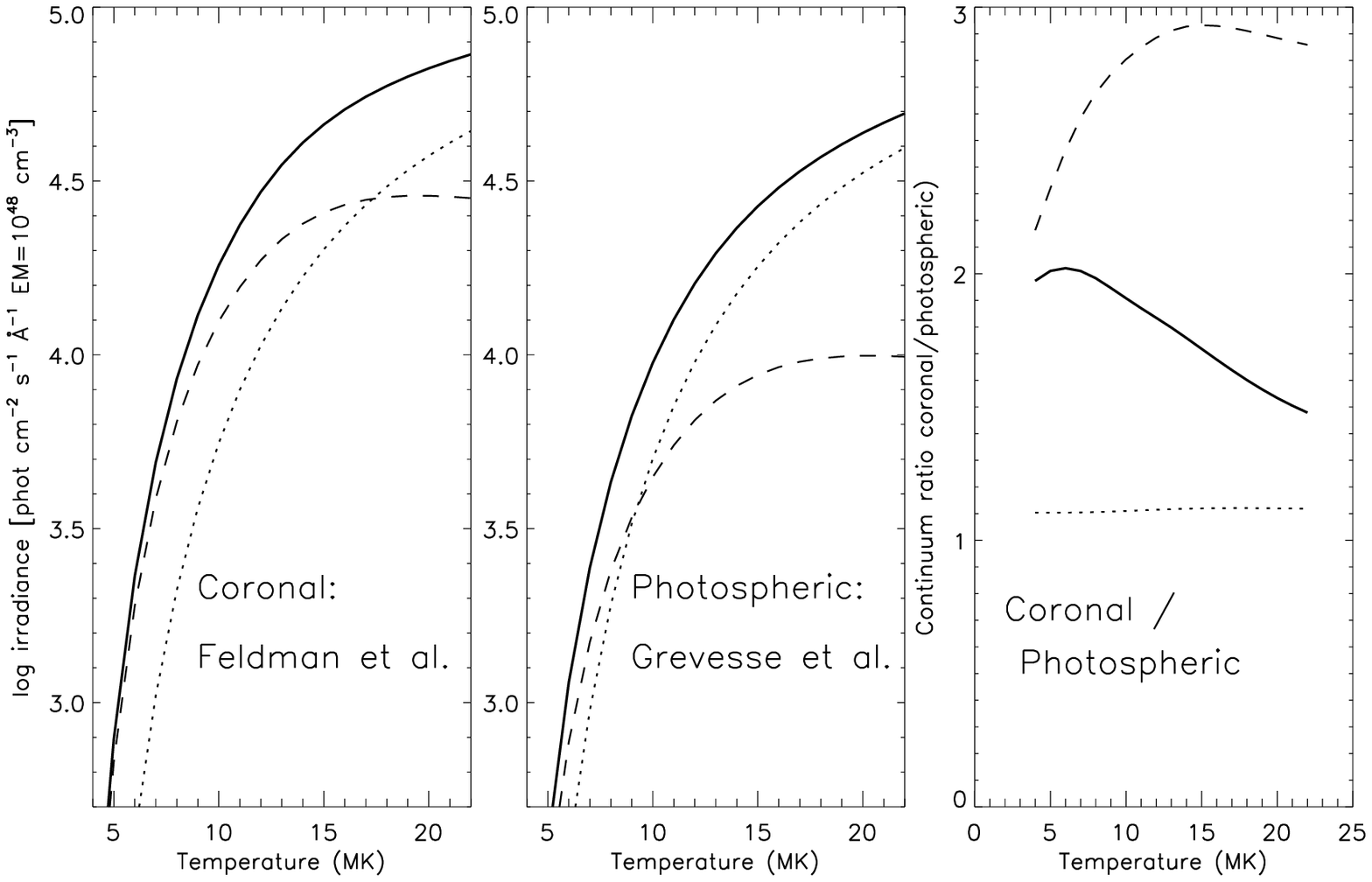}
\caption{Continuum fluxes (irradiances, photon units) at 3.495~\AA\ as a function of $T_e$ calculated from the CHIANTI atomic package. {\it Left panel:} Continua with the \cite{fel92} coronal abundances. {\it Center panel:} Continua with the \cite{gre07} photospheric abundances. {\it Right panel:} Ratio of continuum fluxes calculated with the \cite{fel92}  abundances to those calculated with the \cite{gre07} abundances. The dashed curves refer to free--bound continua, dotted curves to free--free continua, and solid curves to the sum of all continua. Two-photon continua are not shown; they are a factor 30 or more lower than the free--free continuum.  } \label{comparison_cont}
\end{figure}

A further illustration of the effect of abundances is given in Figure~\ref{comparison_cont}: in this case, continuum fluxes at 3.495~\AA\ are plotted against $T_e$ for coronal (\cite{fel92}: left panel)  abundances and photospheric (\cite{gre07}: center panel) abundances and their ratios (right panel). Again, an emission measure of $10^{48}$~cm$^{-3}$ is assumed. There are differences of a factor 2--3 in the free--bound continua, though at longer wavelengths the differences are less. For 3.495~\AA, free--bound emission is equal to free--free at a temperature of 17~MK for coronal abundances, but a much smaller temperature (9~MK) for photospheric abundances.

X-ray continua from the analytical formulae of \cite{culact70} and \cite{gro78}, calculated on the basis of coronal abundances adopted by these authors, have been widely used in the past. These abundances are different from the more definitive work of \cite{fel92}, and partly as a result of this, there are differences of up to 60\% from the {\sc chianti} curves with the \cite{fel92} coronal abundances for the wavelengths considered here. The wavelength dependence of the total emission is, however, very similar in all three cases.

\section{OBSERVATIONS AND ANALYSIS}\label{obs_anal}

A sample of 19 flares, all with few or no interruptions in the time coverage, was selected from the RESIK data base. A total of 2795 spectra was analyzed. Table~\ref{Spectra_list} gives a list of time periods during flares when RESIK observations were taken and analyzed.\footnotemark\  The {\it GOES} class, peak times, and heliographic coordinates of the host active region are given. The spectra were accumulated in data-gathering time intervals (DGIs) that were automatically adjusted using on-board software according to the incident X-ray flux. During large flares of {\it GOES} class equal to at least M1, the DGI typically decreased from a few minutes at the flare onset to only $\sim 2$~s at flare peak, then increased again during the flare decline. The pulse-height analyzers on RESIK (by which solar X-ray photons are distinguished from crystal fluorescence photons) had settings that were varied over the first few months of the spacecraft mission and spectra examined to find the optimum settings; the flare spectra analyzed in this study were taken when the settings were established to be close to optimum. The spectra in channels~1 and 2 were converted to absolute flux units (photons cm$^{-2}$ s$^{-1}$ \AA$^{-1}$) using calibration factors for both the crystal reflectivities and the proportional counter detectors; the procedure is described by \cite{syl05}. The resulting fluxes are estimated to have a $\sim 20$\% accuracy. RESIK was turned off during spacecraft passages through the South Atlantic Anomaly and the auroral ovals near the terrestrial poles (the spacecraft orbital inclination is $82.5^\circ$); at other times, there is a low particle background (between 0.01 and 0.05 counts bin$^{-1}$ s$^{-1}$ for each of the four detectors, evaluated from observed counting rates in ``hidden" or non-solar bins) which was subtracted from the observed counts in each detector.

\footnotetext{The data are
made available to the solar physics community via the World Wide Web site
http://www.cbk.pan.wroc.pl/RESIK\_Level2 based at the Space Research Centre, Wroc{\l}aw, Poland.}

\begin{deluxetable}{lcccc}
\tabletypesize{\scriptsize} \tablecaption{RESIK S{\sc pectra} {\sc with} C{\sc ontinuum} E{\sc mission}
\label{Spectra_list}} \tablewidth{0pt}

\tablehead{\colhead{Date} & \colhead{Time of Flare} &  \colhead{{\it GOES} Class} & \colhead{Location} & \colhead{Number of}
\\
& \colhead{Maximum (UT)} & & &\colhead{Spectra  } }

\startdata

2002 August     3 & 19:07 & X1.0 & S16W87 &  413 \\
2002 September 10 & 14:56 & M2.9 & S10E43 &  235 \\
2002 September 29 & 06:39 & M2.6 & N10E20 &   15 \\
2002 October    4 & 05:38 & M4.0 & S19W09 &  317 \\
2002 November  14 & 22:26 & M1.0 & S15E62 &  128 \\
2002 December   4 & 22:49 & M2.5 & N14E63 &   12 \\
2002 December  17 & 23:08 & C6.1 & S27E01 &  166 \\
2003 January    7 & 23:30 & M4.9 & S11E89 &  113 \\
2003 January    9 & 01:39 & C9.8 & S09W25 &  297 \\
2003 January   21 & 02:28 & C8.1 & N14E09 &   69 \\
2003 January   21 & 02:50 & C4.0 & N14E09 &   25 \\
2003 January   21 & 15:26 & M1.9 & S07E90 &  290 \\
2003 February   1 & 09:05 & M1.2 & S05E90 &  133 \\
2003 February   6 & 02:11 & C3.4 & S16E55 &   34 \\
2003 February  14 & 02:12 & C5.4 & N12W88 &   44 \\
2003 February  14 & 05:26 & C5.6 & N11W85 &   64 \\
2003 February  15 & 08:10 & C4.5 & S10W89 &  299 \\
2003 February  22 & 04:50 & B9.6 & N16W02 &   19 \\
2003 February  22 & 09:29 & C5.8 & N16W05 &   29 \\

\\

\enddata

\end{deluxetable}

The wavelengths chosen for estimating the continuum emission were centered on 3.495~\AA\ (width of range 0.051~\AA) and 3.762~\AA\ (0.028~\AA) for channel~1, and 3.840~\AA\ (0.051~\AA), 4.070~\AA\ (0.011~\AA), and 4.220~\AA\ (0.031~\AA) for channel~2. No significant line emission is known to occur in these ranges according to the {\sc chianti} atomic database and the spectral line list of \cite{kel87}, though the S recombination edge at 3.846~\AA\ mentioned in \S 2 is within the third band. Fluxes in these wavelength ranges were determined for as many time intervals in flares as were available.

In our analysis, we assumed that a single temperature characterizes the emission in the 3.5--4.22~\AA\ range. This is not true for a wider wavelength range, but for spectroscopic purposes this is a good approximation for the narrow ranges considered here. To examine the temperature dependence of continuum fluxes, there are a number of options for finding the temperature. First, several line ratios are available, the most suitable being the ratios of Si~{\sc xii} dielectronic line features to Si~{\sc xiii} parent lines in channel~4 \citep{phi06a}. However, these ratios are appropriate for softer wavelengths (up to 6~\AA) than those discussed here  and are therefore likely to yield lower temperatures. Secondly, the ratio of total emission in RESIK channels 1 and 4, $I(3.40-3.80 {\rm \AA})/I(5.00-6.05 {\rm \AA})$, is sensitive to temperature ($T_{\rm 1/4}$) through the fact that the same Si lines as well as a  Si~{\sc xiv} line are included in channel~4, while continuum and weak K~{\sc xviii} lines are included in channel~1, with different temperature sensitivity. Thirdly, the ratio of the two {\it GOES} channels is a temperature indicator ($T_{\rm GOES}$), this being widely used in previous analyses of flares. It was shown in previous work \citep{phi06a} that $T_{\rm 1/4}$ is linearly related to $T_{\rm GOES}$. However, in the present analysis plots of the observed continuum flux against $T_{\rm GOES}$ were found to have significantly smaller scatter than those plotted against $T_{\rm 1/4}$, particularly when the emission in channel~1 was weak, suggesting that $T_{\rm GOES}$ is a better temperature indicator. We therefore chose to use $T_{\rm GOES}$ in the analysis of the RESIK spectra discussed here. The work of \cite{whi05} was used for the conversion of the {\it GOES} ratio to temperature $T_{\rm GOES}$. In deriving $T_{\rm GOES}$, we chose not to remove a pre-flare level from the {\it GOES} flux ratios since this was not done for the RESIK spectra. It is possible that estimates of $T_{\rm GOES}$ have lower precision for low ($T_{\rm GOES} \lesssim 5$~MK) since for such temperatures the signal in the higher-energy {\it GOES} channel is relatively weak.

\begin{figure}
\epsscale{1.}
\plotone{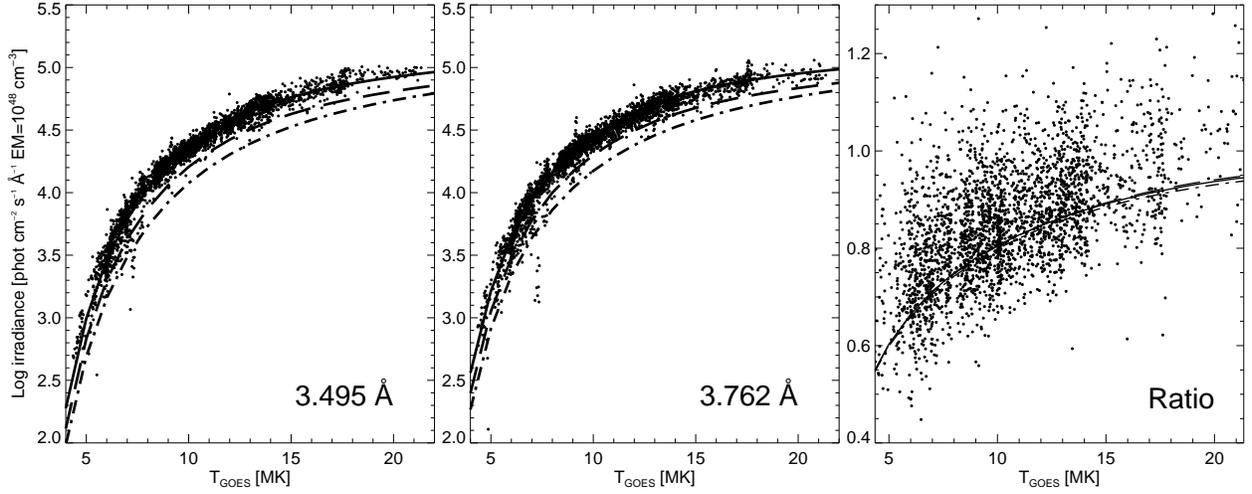}
\caption{Measured continuum fluxes (points) in RESIK channel~1 at 3.495~\AA\ ({\it left panel}) and 3.762~\AA\ bands ({\it center}) plotted on a logarithmic scale against the temperature $T_{GOES}$ determined from {\it GOES} ratios, all normalized to an emission measure of $10^{48}$~cm$^{-3}$ (emission measure estimated from {\it GOES}). These are compared with calculated continua (free--free plus free--bound) from {\sc chianti} with the coronal abundances of \cite{fel92} (solid line), of \cite{flu99} (dashed line), and with the photospheric abundances of \cite{gre07} (dot--dash line).  The ratio of the 3.495~\AA\ and 3.762~\AA\ continuum fluxes are shown in the right panel, with corresponding line styles.  } \label{ch1_cont}
\end{figure}

\begin{figure}
\epsscale{1.}
\plotone{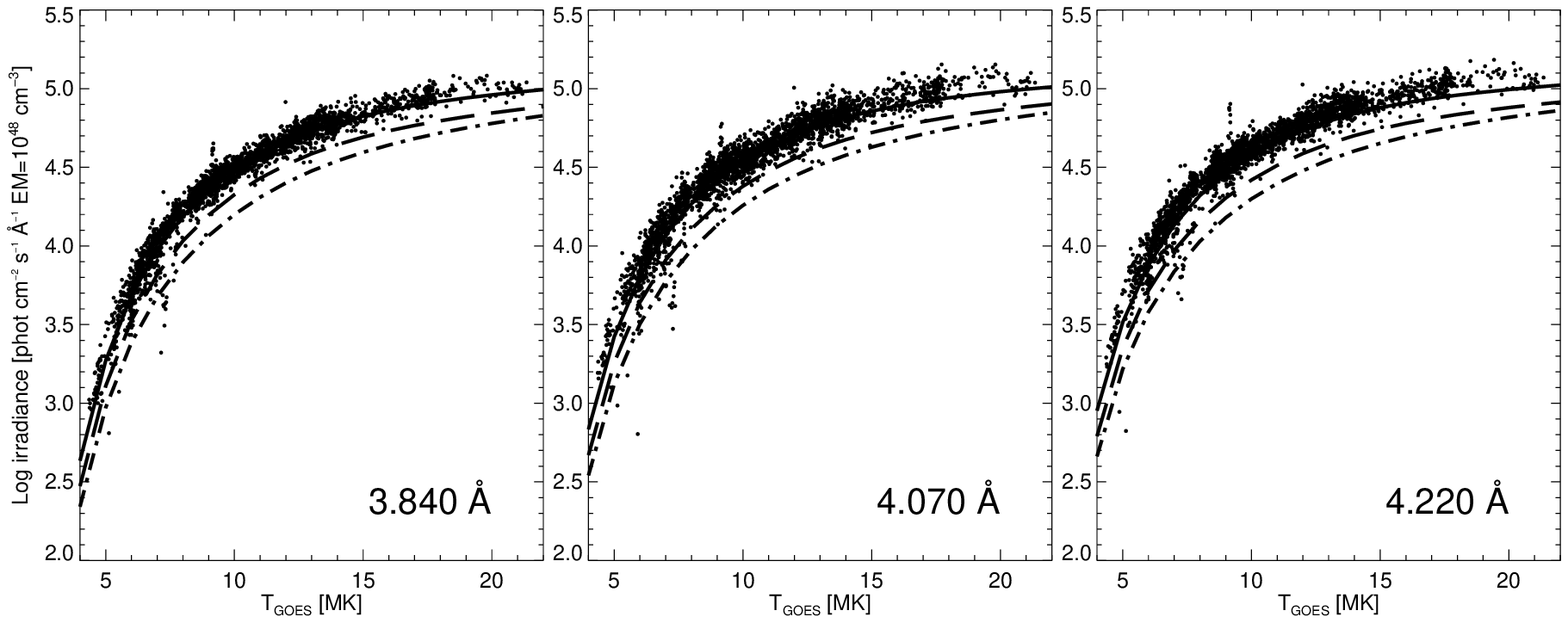}
\caption{Measured continuum fluxes in RESIK channel~2 at 3.840~\AA\ ({\it left panel}), 4.070~\AA\ ({\it center}), and 4.220~\AA\ bands ({\it right}) plotted on a logarithmic scale against $T_{GOES}$, normalized to an emission measure of $10^{48}$~cm$^{-3}$. Calculated continua with the coronal abundances of \cite{fel92} (solid line), \cite{flu99} (dashed line), and photospheric abundances of \cite{gre07} (dot--dash line).  } \label{ch2_cont}
\end{figure}

The plots of all continuum flux measurements against $T_{\rm GOES}$ are shown for the two continuum bands of channel~1 in Figure~\ref{ch1_cont}, together with the flux ratio of the two bands, and the three continuum bands of channel~2 in Figure~\ref{ch2_cont}. The logarithm of the flux is plotted, with the fluxes normalized to an emission measure of $10^{48}$~cm$^{-3}$. Each point represents an observed level of continuum in a time interval during the flares given in Table~\ref{Spectra_list}, with the temperature and emission measure estimated from the {\it GOES} ratio averaged over the time of each RESIK measurement. The total range of $T_{\rm GOES}$ is from 4~MK to 22~MK. The observed points are compared with calculated continua from the {\sc chianti} code with coronal abundances, shown as solid lines (abundance set of \cite{fel92}) and dashed lines \citep{flu99}, and photospheric abundances \citep{gre07}, shown as dotted lines.  The RESIK points agree significantly better with the {\sc chianti} curves assuming the \cite{fel92} coronal abundances for the continuum bands 3.495~\AA\ and 3.762~\AA, with the observed points occurring at higher fluxes than are predicted by the coronal abundances of \cite{flu99} or the photospheric abundances of \cite{gre07}. The scatter of the points around the curve with the \cite{fel92} abundances is approximately 0.1 in the logarithm (26\%). The standard deviation of the points is therefore less than this amount, in agreement with the estimated uncertainty in the RESIK absolute calibration ($\pm 20$\%).  The observed ratios of these two bands (plotted in the right panel of Figure~\ref{ch1_cont}) are consistent with all three abundance sets. For the continuum fluxes in channel~2 (Figure~\ref{ch2_cont}), there is also agreement with the theoretical curves assuming the \cite{fel92} abundances to within measurement uncertainties for temperatures up to about 15~MK, above which the points are up to about 40\% larger, slightly more than the estimated uncertainties. This departure may arise through the fact that a single temperature given by $T_{\rm GOES}$ less accurately describes the temperature of the emission: possibly a second component with different temperature is needed for these cases. Nearly all the observed points are above the other curves, by about 0.2 in the logarithm (60\%) in the case of the \cite{flu99} abundance set and about 0.3 in the logarithm (a factor 2) in the case of the \cite{gre07} abundance set.

There are important implications for coronal element abundances in the curves of Figures~\ref{ch1_cont} and \ref{ch2_cont}. The agreement of the RESIK continuum fluxes with the {\sc chianti} curves with the \cite{fel92} abundances for the wavelength bands centered on 3.495~\AA, 3.762~\AA, 3.840~\AA, 4.070~\AA, and 4.220~\AA\ implies that the abundances of elements giving rise to the largest contributions in these ranges, viz. Si, Fe, Mg, and O, are at least approximately correct. The \cite{fel92} abundances of Si, Fe, and Mg are approximately a factor 2 more than those of \cite{flu99}, and the abundance of O 40\% more. Thus, an abundance of Fe that is a factor 4 more than the photospheric \citep{gre07} is consistent with these results rather than one that is only a factor 2 more, as with the \cite{flu99} abundances. This is in agreement with Fe/H abundance results from an analysis of {\it RHESSI} thermal flare spectra \citep{phi06b}. Out of 27 flares included in this analysis, 19 flares had an Fe/H abundance within 20\% of the \cite{fel92} value; of the remaining eight flares, four had an Fe/H abundance more than 20\% different from the \cite{fel92} value, while four flares had an equivocal result.

There are some points falling below all three theoretical curves, which are mostly those for the flare of 2003 February~22 (maximum time 09:29 U.T.). They are measured continuum fluxes for the initial stages of this flare, when it appears that an isothermal approximation for the emission is not a valid assumption as with other points in later stages of flares. Also, although there seems to be no significant departure from the curves for points in individual flares, indicating possible abundance variations during flares, we are investigating this in work in progress, in which line emission from particular elements is being used to evaluate abundances of some elements.

\section{CONCLUSIONS}\label{concl}

We have reported on measurements with the RESIK instrument of X-ray continuum emission at five line-free regions (3.495, 3.762, 3.840, 4.070, 4.220~\AA) in a total of 2795 spectra in 19 solar flares. With temperature and emission measure determined from the ratio of emission in the two {\it GOES} channels, the RESIK continuum measurements, normalized to an emission measure of $10^{48}$~cm$^{-3}$, plotted against temperature of the observed emission  closely follows the theoretical free--free and free-bound continuum using the {\sc chianti} atomic code with coronal abundances from \cite{fel92}. The continuum measurements are about 60\% higher than the curves for the \cite{flu99} abundances, and  are higher by about a factor 2 than the summed free--free and free--bound continua calculated with the photospheric abundances of \cite{gre07}. Thus the observed continuum in these spectral regions is consistent with the coronal abundances of \cite{fel92}, suggesting that the abundances of those elements (O, Si, Mg, Fe) which are large contributors to the free--bound continuum are about four times photospheric for this sample of flares. Apart from points taken during the initial stages of the flare of 2003 February~22 (maximum 09:29 U.T.), an isothermal plasma assumed in this work appears to be justified in the narrow wavelength bands studied here. In work in progress, in which RESIK flare continua are compared with those at other wavelengths from {\it RHESSI}, the emitting plasma will be taken to have a multi-temperature structure.

\acknowledgments

We are grateful for financial help from  the European Commission's Seventh Framework Programme (FP7/2007-2013)
under grant agreement No. 218816 (SOTERIA project, www.soteria-space.eu), the Polish Ministry of
Education and Science Grant N N203 381736,  and  the UK--Royal Society/Polish Academy of Sciences International Joint Project (grant number 2006/R3) for travel support. {\sc chianti} is a collaborative project involving Naval Research Laboratory (USA), the Universities of Florence (Italy) and Cambridge (UK), and George Mason University (USA).

\end{document}